\documentclass[preprint,a4paper,11pt]{article}
\usepackage[utf8]{inputenc}
\usepackage[super]{natbib}
\usepackage{amsmath}
\usepackage{url}
\usepackage{hyperref}
\usepackage{epsfig}
\usepackage{qtree}
\usepackage {tikz}
\usetikzlibrary{calc,trees,positioning,arrows,chains,shapes,decorations.pathreplacing,decorations.pathmorphing,matrix}
\usepackage{tikz-uml}
\tikzstyle{rect} = [rectangle, draw, fill=white!20, node distance=2cm, text width=6em, text centered, sharp corners, minimum height=3em]
\usetikzlibrary {positioning,arrows,shapes}
\usetikzlibrary{plotmarks}
\usepackage{listings}
\usepackage{makecell}
\usepackage{color}
\usepackage{algorithm}
\usepackage{algorithmic}
\usepackage{amssymb}

\usepackage{multirow}
\usepackage{enumitem}
\usepackage{soul}
\setlist[description]{leftmargin=\parindent,labelindent=\parindent}
\usepackage{doi}
\lstdefinelanguage{json}{
    basicstyle=\footnotesize\ttfamily,
    numberstyle=\scriptsize,
    stepnumber=1,
    numbersep=8pt,
    showstringspaces=false,
    breaklines=true
   }
   
\newcommand{\funcname}[1]{{\operatorname{#1}}}  

\newcommand{\organizedb}{organized behavior}
\newcommand{\organicb}{organic behavior}

\newcommand{\tracedHashtag}{\textit{tracedHT}}

\newcommand{\userFeatures}{user features}
\newcommand{\seedTweets}{\textit{seed tweets}}
\newcommand{\ST}{\textit{ST}}
\newcommand{\expandedTweetSet}{\textit{expanded tweet set}}
\newcommand{\formulaspace}{\;}

\def\kbbox[#1,#2,#3,#4,#5]#6{
        \draw[dashed] node[draw,color=gray!50,minimum
        height=#1,minimum width=#2] (#4) at #5 {}; 
        \node[anchor=#3,inner sep=2pt] at (#4.#3)  {#6};
        }
        
\def\myarm{1cm}
\def\myangle{0}
\tikzset{
  arm/.default=1cm,
  arm/.code={\def\myarm{#1}}, 
  angle/.default=0,
  angle/.code={\def\myangle{#1}} 
}

\tikzset{
    myncbar/.style = {to path={
        let
            \p1=($(\tikztotarget)+(\myangle:\myarm)$)
        in
            -- ++(\myangle:\myarm) coordinate (tmp)
            -- ($(\tikztotarget)!(tmp)!(\p1)$)
            -- (\tikztotarget)\tikztonodes
    }}
}

\newcommand\ceil[1]{\Bigg\lceil#1\Bigg\rceil}

\begin{document}

\title{Organized Behavior Classification of Tweet Sets using Supervised Learning Methods}

\author{Be{\u{g}}enilmi{\c{s}}, E.\thanks{Corresponding author. MS, Big Data Architect, OREDATA, Istanbul, Turkey. erdem.begenilmis@gmail.com. This work was done as part of his MS thesis in the Department of Computer Engineering, Bogazici University, Istanbul, Turkey.} and \"{U}sk\"{u}darl\i, S.\thanks{PhD, Department of Computer Engineering and Complex Systems Research Laboratory, Bogazici University, Istanbul, Turkey.}} 

\date{}
\maketitle

\begin{abstract}
During the 2016 US elections Twitter experienced unprecedented levels of propaganda and fake news through the collaboration of bots and hired persons, the ramifications of which are still being debated.
This work proposes an approach to identify the presence of organized behavior in tweets.
The Random Forest, Support Vector Machine, and Logistic Regression algorithms are each used to train a model with a data set of 850 records consisting of 299 features extracted from tweets gathered during the 2016 US presidential election. 
The features represent user and temporal synchronization  characteristics to capture coordinated behavior.
These models are trained to classify tweet sets among the categories: organic vs organized, political vs non-political, and pro-Trump vs pro-Hillary vs neither.
The random forest algorithm performs better with greater than 95\% average accuracy and f-measure scores for each category.
The most valuable features for classification are identified as user based features, with media use and marking tweets as favorite to be the most dominant.
\end{abstract}


\section{Introduction}
\label{sec:introduction}

Twitter is one of the most popular microblogging platforms with nearly 328 million users~\footnote{https://www.statista.com/statistics/282087/number-of-monthly-active-twitter-users(April 2017)}. 
In recent years, Twitter has become an alternative platform to main stream media for getting news. 
Politics is one of the most prominent uses of social media platforms due to their facilitation of reaching the masses. 
Numerous guides and tips are disseminated for political campaigns for the effective use of Twitter\footnote{https://www.onlinecandidate.com/articles/political-candidate-twitter-tips.}
Most politicians have a verified Twitter accounts, which they use to directly communicate with the public. 

Twitter provides a lucrative platform for diffusing (mis)information at a very low cost \cite{crossChannelMediaCosts}, which can be easily utilized for targeting and manipulating users.  
The 2016 US presidential election, where Hillary Clinton and Donald Trump were presidential candidates, demonstrated the effectiveness of using Twitter very well\cite{FM7090,Ferrara:2016:RSB:2963119.2818717}. 
During this campaign approximately 400,000 social bots (automated agents) generated around 3.8 million tweets corresponding to 19 percent of all campaign related posts \cite{PresedentialElectionOnline}. 

Furthermore, recently Fake News has gained much attention due to its prevalence and persuasive capacity\cite{allcott2017social}.
The impact of believing such news can be dramatic. 
For example, during US election 2016,  a news, which states 
Hillary Clinton used a pizza restaurant, as a front for a pedophile sex ring became a viral, and resulted in an arrest of a person who entered the restaurant with an assault rifle~\cite{pizzaGate2,pizzaGate1}. 
During the 2016 US election, there were numerous fake news such as ``Donald Trump Protester Speaks Out: I Was Paid \$3,500 To Protest Trump's Rally."~\cite{fakeNewsProtestors}.

In order to propagate messages in accordance with an agenda, there is an increasing incidence of employing social media users \cite{BotsStrongerInBrexit}.
An investigation of fake news sites during the 2016 US election revealed that Veles (a Macedonian town with a population of 45,000) was the source of many pro-Trump fake news \cite{fakeNewsMacedonia2,fakeNewsMacedonia1,fakeNewsMacedonia3}. 
In order to rapidly produce fake news, hundreds of teenagers were employed in the city.
Similarly, massive presence of bot accounts in pro-Trump and pro-Hillary messages~\cite{BotsAndAutomationOverTwitter} suggests existence of an organized effort to disseminate a message.

Propaganda calls for coordination and organization. 
When carried out effectively it can be quite deceiving and manipulate opinions and decision.
It can sway election results and impact the future of a country or the world.  
Within collaborative and organized use, terrorist groups and extremists may harm security and peace.
For example, ISIS\footnote{ISIS is an extremist terrorist organization}
and White Supremacy Extremists (FBI declares as ``Domestic Extremist Ideologies''\footnote{FBI known violent extreme groups: \url{https://www.fbi.gov/cve508/teen-website/what-are-known-violent-extremist-groups}.}\enlargethispage{-\baselineskip}) strategically use social media
to recruit new members and to disseminate their propaganda \cite{NazisVsISISTwitter,ISISTwitterCensus}.

The automatic detection of propaganda on social media may help to protect social media users and services from malicious activity. 
This work proposes an approach for the automatic detection of \organizedb, which is likely to be observed in online propaganda.
The term \organizedb\ is  used to refer to a collaborative and coordinated posting behavior involving multiple users, who serve an agenda.
On the other hand, tweets that are posted spontaneously without any predetermined agenda are referred to \organicb\footnote{The organic term is borrowed from \cite{OrganicorOrganizedUrlCao}}\enlargethispage{-\baselineskip}. 

For the automatic detection of  \organizedb, we propose a supervised learning method that utilizes features of collective behavior
in hashtag based tweet sets, which are collected by querying hashtags of interest.
Numerous user and temporal features are extracted from the collected tweets to feed machine learning algorithms to train a model for detecting \organizedb.
After the analysis of over 200 million tweets which are mainly posted during the 2016 US presidential election
The model is trained using a training data set with 851 records.
The 851 records in the training data set reflects the analysis of over 200 million tweets which are posted mainly during the 2016 US presidential election.

All of the features in the training data set are extracted by analyzing
user characteristics \& temporal tweeting patterns, which are independent of the content (except for sentiment based feature) and graph related features.
In other words, they are based on the characteristics of a poster and the presence of temporal coordination among the tweets.
This is akin to sensing that \textit{something is up} without knowing \textit{what is up}.
To discover the latter further analysis would be required on content.
The content and graph related features are not used, because  extraction of these features are not cheap and easy compared to user \& temporal features.
Thus, this might affect the performance of near real time automatic detection.
Given the massive quantity of content, such detection is deemed to be of significance, 
since malicious content can quickly become trending and spread to millions of users. 

Finally, the extracted features are used to train models to predict tweet set behavior in each of the three categories [\organizedb\ vs. \organicb], [pro-Trump vs. pro-Hillary vs. none] and  [political vs. non-political].
The trained models gave promising results for each category, where the Random Forest algorithm consistently resulted in high scores with an average accuracy and f-measure greater than $0.95$. 
Although the training data size, 851 records, is somewhat small for a supervised learning approach, the resulting classification models are able to distinguish tweet sets among the categories and are encouraging for use in larger systems.

This work proposes a proof of concept method for the near real time detection of propaganda and \organizedb.
The main contributions of this work are as follows:  a classification model to detect organized behavior in tweet sets based on only user and temporal posting features, a prototype implementation for feature extraction \& classification, qualitative and quantitative description of features of tweet sets involving the controversial 2016 US election.

The remainder of the paper is structured as follows:
Section~\ref{sec:background} describes background information required to follow this work; 
Section~\ref{sec:relatedwork} presents related work;
Section~\ref{sec:approach} describes the overall approach; 
Section~\ref{sec:model} presents the proposed model; 
Section~\ref{sec:features} provides definitions of the extracted features; 
Section~\ref{sec:experiment} presents the experiments and results; 
Section~\ref{sec:discussion} discusses the results and future work;
Section~\ref{sec:futureworkandconclusions} concluding remarks.

\section{BACKGROUND}
\label{sec:background}

This section provides information about the tools and techniques utilized during the prototype development of the model.

\subsection{Twitter}
\label{sec:twitter}

There are many Twitter related terms\cite{twitterGlossary}, the most relevant of which are described here. 
Twitter is a social networking application that supports posting short messages called \textit{tweet}s.
The main connection among users is the \textit{follow} relations. 

The act of propagating a tweet is called \textit{retweet}ing.
The more a tweet is retweeted the more visible it becomes.
However, since the volume of tweets is very high, the chances of an individual tweet being seen is very low.

A reference to another user is made using  a \textit{mention}\footnote{For more information about Twitter entities see: \url{https://developer.twitter.com/en/docs/tweets/data-dictionary/overview/entities-object1}} , which is a user name preceded by the \@ character (i.e. @theRealDonaldTrump). 
A \textit{reply} to a tweet, is a tweet whose first token is a mention corresponding to the replier.
Twitter sends notifications to users when they are mentioned, bringing attention to who did the mentioning and why.

A \textit{hashtag}, a string preceeded by \#, is used to tag a tweet. 
When used in multiple tweets by multiple people, it serves to  create a relationship among tweets.
This is widely performed during events as well as with people and topics.  

The Twitter REST API\footnote{Twitter REST API: \url{https://developer.twitter.com/en/docs} (accessed:2017-09-28).} provides a great variety of methods with some access limits to prevent excessive queries \footnote{For rate limits see:\url{https://developer.twitter.com/en/docs/basics/rate-limits}.}
In this study, the User Timeline API is used for fetching the tweets of a user in a given time interval. 

The Twitter Search API  supports queries subject to various criteria, such as a hashtag, or a geographic location.
Queries return results from a sampling of tweets which are posted in recent 7 days.
Results of queries, are therefore, subject to the limitations of the sampling algorithms of the Twitter API, which claims to be optimized for relevance.

\subsection{Supervised Classification Methods}
\label{sec:ClassificationMethods}

In machine learning, classification refers to the task of automatically identifying the category of an entity.
There are three main approaches to learning: supervised, semi-supervised, and unsupervised\cite{Alpaydin:2010:IML:1734076}.
This work relies on supervised learning methods where a model is trained with datasets consisting of labeled data samples (observations) that identify the class of the data. 
The models generated with these approaches are used to predict the classes in unlabeled datasets. 
This work examines three supervised learning methods to detect organized behavior: random forest, support vector machine (SVM), and logistic regression. 

The random forest approach creates a forest of random decorrelated decision trees from a trained feature set \cite{PracticalDataScienceWithR}.
The resulting forest is used to predict a classification based on the most predicted class by its decision trees. 
The support vector machine method aims to identify a hyperplane that best divides a dataset into two classes.
It finds hyperplanes in high dimensional spaces with the greatest possible margin between it and the data points, which increases the confidence\cite{Alpaydin:2010:IML:1734076,BurgesSupportVectorMachines}.
SVMs work better on smaller and cleaner datasets.
Logistic regression is a statistical approach that is applied to machine learning\cite{PracticalDataScienceWithR}.  
It is well suited for binary classification problems. 

In supervised learning methods, there is the risk of generating overfitted models when training data sets are small or there are a large number of features. 
An overfitted model results in good accuracy with training data,  but low accuracy with test data. 
The use of resampling techniques\cite{BootstrappingAdvisingResearchMethods} makes it possible to train non-biased classifier models by increasing the variance in the training data set. 
In statistics, resampling methods may be used to generate random subsets to test the success of classification models.
The ten fold cross validation divides the training data set into ten subsamples. 
During the training phase, it generates models from nine subsamples of the training data, which are tested with the tenth subsample. 
This process is repeated ten times for each subsample in training data and test data.
The results of these tests are averaged to determine whether the generated model is satisfactory \cite{CrossValidationkohavi1995study}.

Principal component analysis (PCA) is a statistical method for explaining data with large number of variable using a smaller number of variables called the principal components. 
PCA aims to find the minimum number of uncorrelated features with the highest variances to reduce dimension in the data set~\cite{Alpaydin:2010:IML:1734076}.

The Apache Spark\cite{sparkMlLibguide} machine learning library provides a wide range of support including all the classifications methods described here. 
In the prototype implementation, classification experiments have relied on Spark while PCA and feature selection processes are done by Weka\cite{wekaDefinitions}, which is a data mining software.

\section{Related Work}
\label{sec:relatedwork}

The work in the area of detecting online behavior can be categorized into three main types:  behavior detection, spam detection, content and group detection. 

Cao \cite{OrganicorOrganizedUrlCao} examines the URL sharing behavior in Twitter and distinguishes the sharing behaviors as organic and organized.
For the detection of organized urls, a graph is generated with nodes representing users and edges representing the use of the same URLs in posts. 
A community detection algorithm using the Louvain method\cite{comlargenet} is applied to the graph to extract user groups. Then, URL and posting time based features of those users are extracted and supervised classifications methods, such as Random Forest, NBTree, SMO and LogitBoost, are applied on manually labeled training data with 406 organic and 406 organized records.
After the training phase with 10-fold cross validation, it is shown that applied approach gives a performance around 0.8, in all methods, while Random Forest works best with high F measure (0.836) and ROC area (0.921).

Similarly, Ratkiewicz \cite{DetectingAndTrackingPoliticalAbuse} studied astroturf political campaigns, which are run by politically-motivated individuals and organizations that use multiple centrally-controlled accounts.
During the study, topological, content-based and crowdsourced features of information diffusion networks on Twitter are extracted. 
These features are mainly extracted from composed directed graph whose nodes represent individual user accounts and edges represent the retweet, mention, reply events between users. 
Supervised Learning Methods are used for understanding early stages of viral diffusion. It is presented that their approach gives accuracy results better than 96\% in the detection of astroturf content in a dataset of 2010 US midterm elections. 

In another study about the classification of group behaviors \cite{ClassificationofGroupBehaviors}, it is aimed to detect criminal and anti-social activities in social media.
To detect these behaviors, the theories of group behaviors and interactions were developed. 
Also, graph matching algorithms are applied to explore consistent social interactions, which suggest that complex collaborative behaviors can be modeled and detected using a concept of group behavior grammars.

Also, in order to disclose spam URLs, Cao \cite{DetectingSpamURLCao} analyzed behavioral features in three categories : click-based, posting based, clicking statistics. 
With these features, a training data set is created with 1,049 spam and benign urls by checking the labels of urls in the tweets from a category website.
Another training data set is created by manual labelling of 219 benign and 79 spam urls.
For the both behavioral features in two training data sets, random forest algorithm is used as supervised learning method by using 10-fold cross validation.
The algorithm is trained by using  all feature sets together and separately such as click-based, posting based, clicking statistics. With this approach, 86\% of accuracy is found by using all features for the training.

In another research, political bot accounts, who take place in Brexit Referendum and play strategic role in referendum conversations, are analyzed~\cite{BotsStrongerInBrexit}. 
It is found that these bots use excessively the family of hashtags associated with the argument for leaving the European Union, and utilize different levels of automation. 
It is stated that these bots, which compose one percent of sampled accounts, generate almost a third of all the messages for the Brexit referendum contents in Twitter.

When Cao's study about organized urls \cite{OrganicorOrganizedUrlCao} is compared with our study, it will be seen that the main focus of this study \cite{OrganicorOrganizedUrlCao} is to detect organized URL behaviors, while our study aims to detect organized behaviors.
Also, the extracted and derived features are different from each other.
However, in both study, same classification method approaches are used and random forest algorithm gave the most promising results.
Furthermore, when  Ratkiewicz's study about astroturf political campaigns
\cite{DetectingAndTrackingPoliticalAbuse} is compared with our study, same supervised classification methods are used in both studies, while
feature extraction phases are different from each other.
In their study,  features are extracted from a constructed graph, while in our study they are extracted from users and temporal tweets.
Also in the study about group classification \cite{ClassificationofGroupBehaviors}, group behaviors are identified as a result of graph matching algorithms. In our study no graph related algorithm is used nor a graph related feature is extracted due to performance concerns for real time detection.

In the work about \cite{DetectingSpamURLCao}  identifying spam urls, 
similar works have been done considering feature extraction and classification methods.
However, our work focuses on detection of organized behavior patterns and features of user \& temporal tweets. 

There are also studies about content and group detection which are in parallel with our study \cite{AutoDetectionElectionRelatedTweets,DetectingJihadistMessages,AutomaticClassificationMuseumTweets}. In all studies either underlying content or group is detected. However, in all of them the detection is done by using language or topic related features. On the other hand, in our study this detection has been made with features which do not contain network and content related features except the sentiment features.

\section{Organized Behavior Characteristics}
\label{sec:approach}

In Twitter approximately $6,000$ tweets are posted per second \footnote{http://www.internetlivestats.com/twitter-statistics}.
With such huge numbers of tweets and users, getting messages seen is not a trivial task.
In order to increase the likelihood of the message's observation,  some strategies are required.

To gain insight about how organized behavior is manifested in Twitter, specifically in the political domain,
we manually inspected tweets with chosen top trend hashtags.
These hashtags are selected as candidate organized hashtags, if they seemed suspicious after reading some of their tweets. 

To get  larger numbers of such tweets, Twitter Search API is used to fetch tweet sets for each candidate hashtag.
To gain an overall impression, the following values are computed for each tweet set: 
the percentage of distinct words ($DW(\%)$), 
the average tweet count per user ($TPU(\mu)$), 
the percentage of retweets($RT(\%)$), and 
the variance and standard deviation of hashtags ($HT(\sigma^2)$ and $HT(\sigma)$).
This enabled a comparison among tweet sets, which are shown in Table~\ref{table:manuelInspectionValues}.

A low value of $DW(\%)$ indicates a lack of diversity in vocabulary,
while a high value of $TPU(\mu)$ shows that the users who posted in the collection tend to repeatedly post the same hashtag. 
Also, high values of $HT(\sigma^2)$ \& $HT(\sigma)$ indicate the use of multiple hashtags in the collection.
This can be observed in the so-called viral activity, where multiple hashtags are used to reach as many users as possible.
Similarly, high $RT(\%)$ values indicate less original content. 

In addition to the statistical computations, for each user in the analyzed tweet set, 
other tweet sets with similar hashtag are analyzed to see the user's participation with other hashtag also.
In this step, two hashtags are thought as similar
if manual inspection of their tweets \& hashtag names reveals that 
they have a common point based on content and targeted audience.
The amount of users, who exist in multiple tweet set, also considered during manual inspection of 
organized tweet sets(see~Table~\ref{table:sameUserAnalysis}) . 

\begin{table}[htbp]
	\vskip\baselineskip 
	\caption[Features inspected during manual inspection of collections.]{Tweet sets characteristics inspected during manual inspection.}
	\begin{center}
		\begin{tabular}{|l|r|r|r|r|r|r|}\hline			
			\multicolumn{1}{|c|}{\bfseries Hashtag} & \multicolumn{1}{c|}{\bfseries tweets}  & \multicolumn{1}{c|}{\bfseries DW} & \multicolumn{1}{c|}{\bfseries TPU} & \multicolumn{1}{c|}{\bfseries RT} & 	\multicolumn{1}{c|}{\bfseries HT} & \multicolumn{1}{c|}{\bfseries HT} \\
			 		\multicolumn{1}{|c|}{\bfseries }		& \multicolumn{1}{c|}{\bfseries $(\#)$}  & \multicolumn{1}{c|}{\bfseries $(\%)$} & \multicolumn{1}{c|}{$(\mu)$} & \multicolumn{1}{c|}{\bfseries $(\%)$} & \multicolumn{1}{c|}{\bfseries ($\sigma^{2}$)} & \multicolumn{1}{c|}{\bfseries ($\sigma$)} \\\hline
			\#podestaemails15 & 17,890 & 4.62 & 1.89 & 92.28 & 2.40 & 1.54 \\\hline
			\#BoycottHamilton & 56,523 & 3.92 & 1.54 & 2.12 & 2.72 & 1.65\\\hline
			\#StrongerTogether & 13,581 & 11.9 & 1.64 & 77.06  & 0.46 & 0.68 \\\hline
			\#unitedairlines & 54,506 & 8.13 & 1.28 & 71.86 & 0.75 & 0.86 \\\hline
			\#womansDay & 416,350 & 7.57 & 1.38 & 78.76 & 0.13 & 0.36 \\\hline
		\end{tabular}
		\label{table:manuelInspectionValues}
	\end{center}
\end{table}

The manual inspection of these tweet sets (especially the ones about political propaganda) revealed that the \organizedb\ characteristics 
can be summarized as: sharing a common goal, temporal synchronization among users, and the dissemination of messages.
Similar characteristics are reported in studies about digital and social media activism\cite{CanadianAdvocacy2.0,TheseDaysWillNeverBeForgotten,DigitalActivismNonViolentConflict}.
Basically, to make an impact on Twitter, the message has to get seen and spread. 
The more it is seen the more it can spread. 
The more it spreads the more it persists and is likelier to be seen.
In order to detect \organizedb, we benefit from tweet sets with two criterias: hashtag use and
bot account activity. 

Hashtags are used in organized behaviors~\cite{NazisVsISISTwitter,ISISTwitterCensus}, because
they facilitate a shared context by grouping disjoint tweets.
During the US election 2016, the  \#FeelTheBern hashtag is used by supporters of the democratic candidate Bernie Sanders~\cite{feelTheBern}.
Twitter's top trend functionality also makes hashtag use preferable for \organizedb\ \cite{NazisVsISISTwitter,ISISTwitterCensus}. 

Furthermore, the prevailing presence of bot existence in a tweet set suggests the presence of \organizedb.
During 2016 US elections, approximately 400,000 bots generated around 3.8 million
tweets\cite{PresedentialElectionOnline}.
Similarly, in the first presidential debate, $1/3$ of 1.8 million pro-Trump posts 
and $1/4$ of 600,000 pro-Hillary posts were generated by bots ~\cite{BotsAndAutomationOverTwitter}.
The degree of these bot tweets and the concentration of their messages on same hashtags~\cite{botNews5} suggest that these bots post to serve a goal.
Likewise, use of pro-Trump bots for spreading fake news~\cite{botNews4} also supports this idea. 
These bot activities include all three characteristics of organized behaviors, which were discovered upon our manual inspection,
such as sharing a common goal, the temporal synchronization of tweets, and the dissemination of messages.

\section{An Organized Behavior Detection Model}
\label{sec:model}

The automated detection of organized behavior on Twitter calls for methods that scale to the data corresponding to  massive numbers of posts and users. 
Machine learning approaches are promising in many classification problems related to complex large datasets of social big data\cite{BELLOORGAZ201645}. 

\begin{figure}
	\begin{center}
		\includegraphics[height= 0.25\paperheight, width=1\columnwidth]{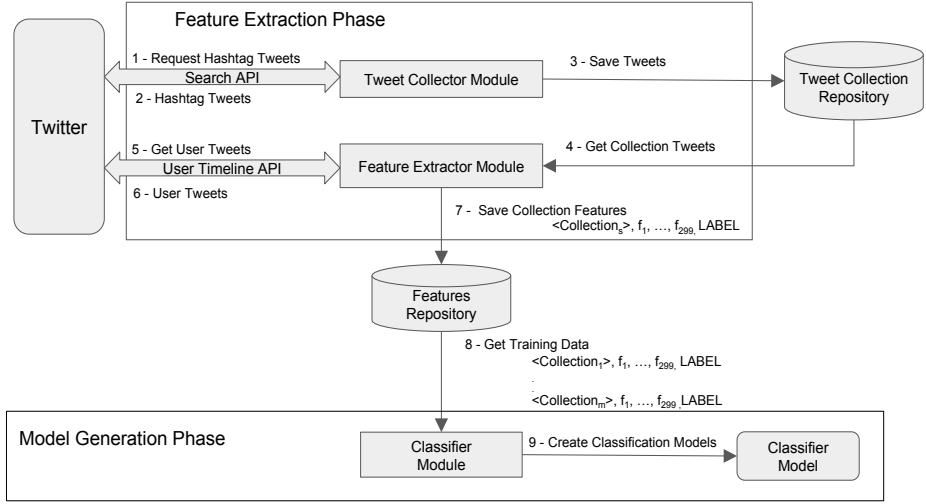}
	\end{center}
	\caption{The overview of generating a classification model for tweet sets.}	
	\vskip\baselineskip
	\label{fig:generalModel}
\end{figure}

This work proposes a simple model based on supervised learning methods. 
Figure~\ref{fig:generalModel} shows the overall approach that consists of two main phases: feature extraction  and model generation.
The feature extraction phase handles collecting the tweets of interest and extracting their features. 
A set of tweets of interest is referred to as a \emph{collection}.
Tweets of interest are chosen to be those that contain a specific hashtag, since hashtags are widely used to increase engagement.
Feature extraction is performed on each collection, resulting in rows of features that are used to train the classifiers.

\subsection{Feature Extraction Phase}
\label{sec:featureExtractionPhase}

\begin{algorithm*}
	\caption[Overview of the feature extraction algorithm.]{\label{alg:GeneralFeatureExtractionAlgorithm}Feature extraction algorithm applied to each tweet set.}
	\begin{center}
				\begin{algorithmic}[1]	
					\STATE \textbf{Hashtag} \textit{tracedHashtag}
					\STATE \textbf{Interval} \textit{analysisInterval}
					\STATE \textbf{Integer} \textit{numDays}
					\STATE \textbf{List} \textit{seedTweets}  
					\STATE \textbf{List} \textit{allUserFeatures} =  $[ ]$
					\STATE \textbf{List} \textit{allTempFeatures} =  $[ ]$
					\STATE \textbf{List} \textit{expandedTweets} = $[ ]$
					
					\STATE $seedTweets \gets getTweets(hashtag=tracedHashtag)$ {\small // Step 1}
					\STATE $users \gets  getUsers(tweets)$	
					\FORALL  {$u \in  users$}	
					\STATE $expandedTweets.add(getTweets(u,numDays,seedTweets)$ {\small// Step 5}
					\ENDFOR
					\FORALL  {$u \in  users$}
					\STATE $userFeatures \gets$ \textit{extractUserFeatures(u,tracedHashtag,} \par
        						\hskip\algorithmicindent \textit{expandedTweets)}
					\STATE $allUserFeatures.add(<u,userFeatures>)$
					\ENDFOR
					\STATE $timeIntervals \gets  getTimeIntervals(tweets,analysisInterval) $
					\FORALL  {$ti \in  timeIntervals$}
					\STATE $tweetsTI \gets getTweets(expandedTweets,ti)$
					\STATE $temporalFeatures \gets extractTemporalFeatures(tweetsTI)$
					\STATE $allTempFeatures.add(<u,temporalFeatures>))$ 
					\ENDFOR
					\STATE $featureStats \gets$ \textit{computeFeatureStats(allUserFeatures,}\par
        						\hskip\algorithmicindent \textit{allTempFeatures)}
					\STATE $trainingDataRecord \gets \{$\textit{allUserFeatures,allTempFeatures,}\par
        						\hskip\algorithmicindent  \textit{featureStats}$\}$
					\RETURN $trainingDataRecord$ {\small // Step 7}
				\end{algorithmic}
	\end{center}
\end{algorithm*}

The feature extraction phase performs two main tasks:  preparation of a collection and  extracting its features (Algorithm~\ref{alg:GeneralFeatureExtractionAlgorithm}). 

A collection is created by fetching tweets with occurrences of a \tracedHashtag\ (hashtag of interest).
Additional tweets of the users who posted these tweets are fetched to expand the collection in order to get more information about them.
The resulting collection is referred to with a hashtag (i.e. \#lockHerUp collection).
A \tracedHashtag\ may be associated with multiple collections if they are created at different times that it is used (i.e. \#guncontrol).

Collection expansion consists of fetching additional tweets that were posted within a given time before and after the time of a post captured in \seedTweets\  (line~11~in~Algorithm~\ref{alg:GeneralFeatureExtractionAlgorithm}).
In our experiments we chose this duration to be one week.
The intent is to capture whether there is a significant difference in the behavior of a user before and after the time of their post in \seedTweets.
The collection of \seedTweets\ expanded with the user tweets is referred to as the \expandedTweetSet (\textit{ET}).
An unqualified reference to a collection should be interpreted as the \expandedTweetSet.
A reference to the \emph{users in a collection} should be interpreted as the users who contributed to the posts in that collection.
Expanding the collection does have a considerable impact on the size of the collection. Table~\ref{table:trainingDataSize} shows size comparisons of a few sample collections.  

All the user and temporal features, as defined in Section~\ref{sec:features}, are extracted for each collection and stored in the \emph{Features Repository}.
User features are computed based on individual characteristics as well as those that compare a user with others in the collection.
The temporal features of tweets are computed based on time intervals corresponding to the time period of the collection (Section~\ref{sec:temporal-features}). 
In our experiments this size was chosen as one hour.
Inspecting the tweets according the time periods attempts to detect the presence of coordinated posts, since organized activities typically strategically schedule posting times.  

Finally, in order to have an overall view of collections, the mean, variance, standard deviation, minimum, and maximum values for all the features are computed and stored in the \textit{Features Repository} as a row of the training data set.

\subsection{Model Generation Phase}

Supervised learning methods are used to train models using the features extracted for the collections. 
Three classifiers are generated using the supervised learning methods of random forest, SVM, and logistic regression.

Furthermore, three kinds of classifications are generated: \organizedb\ vs. \organicb\ ,  \textit{pro-Trump}  vs. \textit{pro-Hillary}, and  \textit{political} vs. \textit{non-political}.
The models were chosen to explore different contexts of organized behavior. 
More specifically, the presence of any kind of organization, a specific kind of group based organization, and topic based organization.

Section~\ref{sec:experiments-results} provides details about the specific collections, the features generated from them, the performance of the generated classifiers, and a comparison of the chosen supervised learning methods.

\section{Feature Selection}
\label{sec:features}

Based on the characteristics of organized behavior and the manual examination of large numbers of tweets (Section~\ref{sec:approach}), two kinds of main features are identified for detecting the presence of organized behavior: \textit{user}  and  \textit{temporal} features.
The user features capture information about the characteristics of users.
The temporal features focus on the timing of tweets (independent of who posted them) to detect the presence of a synchronization.
Recall that, the posting time and synchronization are useful for increasing the visibility of a message.
In other words, user features focus on identifying the participants and the temporal features focus on detecting the presence of organized behavior.

In order to clearly describe the features consider the sets $ST$ (seed tweets, Section~\ref{sec:model}) to be the initial set of tweets fetched with a query for a given hashtag (\tracedHashtag).
Let $ET$ (Expended Tweet Set, Section~\ref{sec:model}), be the set of tweets obtained by expanding $ST$ with additional tweets of users who have posted a tweet $t \in ST$ (see~Figure~\ref{alg:GeneralFeatureExtractionAlgorithm}).

Table~\ref{table:featureFunctions} defines a set of functions used to formulate the features.
Essentially these functions correspond to data fetched using the Twitter API, thus, refer to Twitter specific types like Tweet, Hashtag, and Mention.
Typed sets are denoted with a type name subscripted with $_s$. such as \emph{Hashtag}$_s$ that represents a set of hashtags.

\begin{table}[H]
	\centering
	\caption{Descriptions of user and tweet functions, where $T$ is set of tweets, $t$ is a tweet, $h$ is a hashtag, $u$ is a  user, $D$ is a set of Days, $d$ is a day, $\Delta$ is a duration}
	\label{table:featureFunctions}
\begin{tabular}{|l|l|l|}
\hline
\thead{\textbf{function}}&\thead{\textbf{Type}}&\thead{\textbf{Description}}\\ \hline
\multicolumn{3}{|c|}{\textbf{User functions}}\\ \hline
$\funcname{reg-date}(u)$ & $\mathit{Date}$ & date $u$ registered\\ \hline
$\funcname{following\#}(u)$	& $\mathit{Integer}$ & number of users that $u$ follows\\ \hline
$\funcname{follower\#}(u)$ & $\mathit{Integer}_s$ & number of users who follow $u$\\ \hline
$\funcname{favorite\#}(u)$ & $\mathit{Integer}$ & number of tweets favorited by $u$\\ \hline
$\funcname{tweets}(u,T)$ & $\mathit{Tweet}_s$ & set of tweets posted by $u$ in $T$\\ \hline
$\funcname{tweet\#}(u)$ & $\mathit{Integer}$ & number of tweets posted by $u$\\ \hline
$\funcname{tweetsD}(u,T,D)$ & $\mathit{Tweet{_s}_s}$ & set of tweet sets \\
& & (posted by $u$ on $d \in D$)\\ \hline
$\funcname{users}(T)$ & $\mathit{User}_s$ & set of users in $T$\\ \hline
$\funcname{hashtag?}(u,T,h)$ & $(0|1)$ & $1$ if user has posted a tweet with $h$ in $T$,\\
& & $0$ otherwise\\ \hline

\multicolumn{3}{|c|}{\textbf{Tweet functions}}\\ \hline
$\funcname{hashtag\#}(t)$ & $\mathit{Integer}$ & number of hashtags that occur in $t$ \\ \hline
$\funcname{hashtag?}(t,h)$ & $(0|1)$ &  $1$ if $h \in \funcname{hashtags}(t)$, $0$ otherwise\\ \hline
$\funcname{mention\#}(t)$ & $\mathit{Integer}$ & number of mentions that occur in $t$ \\ \hline
$\funcname{hashtagD}(T,h)$ & $\mathit{Date}_s$ & set of days when tweets with\\
& & $h$ occur in $T$ \\
\hline
$\funcname{media\#}(t)$ & $\mathit{Integer}$   & number of media that occur in $t$\\ \hline
$\funcname{retweeted?}$(t) & $( 0 | 1 )$  & $1$ if $t$ is retweeted, $0$ otherwise\\ \hline
$\funcname{sentiment}(t)$ & \multicolumn{2}{l|}{(\texttt{\small VeryNegative|Negative|Neutral|}} \\
 & \multicolumn{2}{l|}{\texttt{\small Positive|VeryPositive})} \\

\hline
$\funcname{mentions}(t)$ & $\mathit{Mention}_s$  & the set of mentions that occur in $t$\\ \hline
$\funcname{url\#}(t)$ & $\mathit{Integer}$  & number of links (URL) that occur in $t$\\ \hline
$\funcname{temporalTweets}(T,\Delta)$ & $\mathit{Tweet}_s$  & set of temporal tweets in $T$ based on $\Delta$\\ \hline
\end{tabular}
\end{table}

\subsection{User Features}
\label{sec:userFeatures}

As the creators of the tweets that may be part of some collusion, the characteristics of the contributors are expressed with a set of user features. The following features are computed for each user, which corresponds to lines $13-16$ of Algorithm~\ref{alg:GeneralFeatureExtractionAlgorithm}.

\noindent\textbf{Tweet Count}: The total number of tweets posted by a user: $tweet\#(u)$. Used to indicate how active a user is.

\noindent\textbf{Favorited tweet count}: The number of tweets that are marked as \textit{favorite}\footnote{Twitter has renamed Favorite to Like during the final stages of the preparation of this article.}: $favorite\#(u)$. Marking tweets as a favorite is an approach used to gain attention from others to increase follower count (users might follow those who favorite their tweets).

\noindent\textbf{Average tweets/day}: The average number of tweets per day, based on the number of days since the user registered. Used to understand the user's daily tweet frequency, since higher values might indicate the behavior of an automated account.

\begin{equation} \label{eq:avarageTweetsDays}
 \funcname{tweet}\#_{\mu/d}(u) = \frac{tweet\#(u)} {\funcname{today}()-\funcname{regDate}(u)}
\end{equation}

\noindent where $\mathit{today}()$ returns current date and subtraction of days returns the number of days.

\noindent\textbf{Follower degree}: The degree of a user's followers:

\begin{equation} \label{eq:followerDegree}
 \funcname{follower-degree}(u) = \frac{\funcname{follower\#}(u)} {\funcname{follower\#}(u)  +  \funcname{following\#(u)}}
\end{equation}

Follower degrees that approach $1$ indicate a high degree of followers that is typical with popular persons, while degrees approach to $0$ indicate the opposite.
Newly created bots tend to follow numerous users and have very few followers.

\noindent\textbf{Entity use}: Entities are used to relate a tweet to other tweets, users, external resources, or media.
They are used to gain attention, therefore a higher rate would be expected in propaganda.
The use of the entity types: \footnote{For more information about Twitter entities see: \url{https://developer.twitter.com/en/docs/tweets/data-dictionary/overview/entities-object1}}  \textsc{URL}, \textit{Mention}, \textit{Hashtag}, and \textit{Media} are computed for each $u \in \funcname{users}(ET)$:

\begin{subequations} \label{eq:userEntityUse}
\begin{equation}  \label{eq:userHashtagUse}
 \funcname{hastag-use}(u,ET) = \frac{\displaystyle\sum_{t \in tweets(u,ET)} \mathit{hashtag\#}(t)}{\mid tweets(u,ET) \mid}
\end{equation}
\begin{equation} \label{eq:userURLUse}
 \funcname{url-use}(u,ET) = \frac{\displaystyle\sum_{t \in tweets(u,ET)} \mathit{url\#}(t)}{\mid tweets(u,ET) \mid}
\end{equation}
\begin{equation} \label{eq:userMentionUse}
 \funcname{mention-use}(u,ET) = \frac{\displaystyle\sum_{t \in tweets(u,ET)}  \mathit{mention\#}(t)}{\mid tweets(u,ET) \mid}
\end{equation}
\begin{equation} \label{eq:userMediaUse}
 \funcname{media-use}(u,ET) = \frac{\displaystyle\sum_{t \in tweets(u,ET)} \mathit{media\#}(t)}{\mid tweets(u,ET) \mid}
\end{equation}
\end{subequations}

\noindent\textbf{Traced hashtag use}: The number user's tweets that contain the \tracedHashtag\ within ET.
Used to understand how focused the user is to the \tracedHashtag. In \organizedb, it more likely that users concentrated on a hashtag (Section~\ref{sec:approach})
\begin{equation} \label{eq:user-thashtag-use}
\funcname{user-hashtag-use}(u,ET) = \displaystyle\sum_{t \in tweets(u,ET)} \mathit{hashtag?}(t,\tracedHashtag)
\end{equation}

\noindent\textbf{Average daily tweets of \tracedHashtag}: Focuses on the daily use of the \tracedHashtag, considering that a participant in propaganda (or similar activity) would have a persistent use.

\begin{equation} \label{eq:daily-average-tweet}
\funcname{tweet-hashtag\#}_{\mu/d}(u,D,T) = \frac{\displaystyle\sum_{t \in \funcname{tweetsD}(u,T,D)} hashtag?(t,\tracedHashtag)}  {\mid D \mid}
\end{equation}

\noindent  where $T=ET$, $u \in \funcname{users(T)}$, $D = \funcname{hashtagD}(T,\tracedHashtag)$

\noindent\textbf{Average Tweets/Day vs. Average daily tweets of \tracedHashtag}:
This value is used to understand how the user's behavior is in the days when she used \tracedHashtag
compared to her general daily behavior.
If a user posts much more tweets than she normally does for a hashtag, this can be indicator of an \organizedb.

\begin{equation} \label{eq:user-daily-tweet-comparison}
\funcname{user-daily-tweet-comparison}(u,D,ET) = \frac{\funcname{tweet-hashtag\#}_{\mu/d}(u,D,ET)}  {\funcname{tweet}\#_{\mu/d}(u)}
\end{equation}
\noindent\textbf{User creation date:} $userReg(u)$, which can be useful in understanding of collective behavior of users(see~Figure~\ref{fig:sameUserCreationDate}).

\subsection{Temporal Features}
\label{sec:temporal-features}

The temporal aspects of tweets focus on the characteristics of tweets independent of who posted them with the aim of detecting the presence  of a coordinated effort.
As was explained in Section\ref{sec:approach} the dissemination of a message calls for making it visible, which is achieved by synchronous targeted posting to increase the odds of delivery.
Therefore, temporal features of tweets posted within a certain time period are computed and tracked.
Figure~\ref{alg:GeneralFeatureExtractionAlgorithm} shows the algorithm for computing these features.
Unlike with  the \userFeatures\ which consider all of the tweets of users to characterize them, in the case of the temporal features only tweets that contain the \tracedHashtag\ are used.
The initial set of tweets that contain \tracedHashtag\ ($ST$) are extended with tweets posted by $u \in \funcname{users}(ST)$ resulting in $ET$.
The relationship and frequency of users to tweets are expected to be different in coordinated posting behavior.
This section describes the temporal features of tweets posted during an interval $I$,  where:

\begin{equation} \label{eq:temporal-user-daily-tweet-comparison}
T= \bigcup\limits_{\substack{t \in T \\ \funcname{hashtag?}(t,\tracedHashtag) = 1}} \funcname{temporalTweets}(ET,I):
\end{equation}

\noindent\textbf{Entity use}: These are just like the computations made for a user, but in this case the tweet set includes the tweets that were posted during the time interval by many users:

\begin{subequations} \label{eq:EntityUseTI}
\begin{equation}
 \funcname{hashtag-use}(T) = \frac{\displaystyle\sum_{t \in T} \funcname{hashtag\#}(t) \formulaspace} { \mid T \mid}
\end{equation}
\begin{equation}
 \funcname{url-use}(T) = \frac{\displaystyle\sum_{t \in T} url\#(t) \formulaspace  } { \mid T \mid}
\end{equation}
\begin{equation}  \funcname{mention-use}(T) = \frac{\displaystyle\sum_{t \in T}  mention\#(t) \formulaspace } { \mid T \mid}
 \funcname{media-use}(T) = \frac{\displaystyle\sum_{t \in T}  media\#(t) \formulaspace } { \mid T \mid}
\end{equation}
\end{subequations}

\noindent\textbf{Temporal Tweet Per User (TPU)}: This feature is used to
check tweet per user in temporal basis. It is extracted because higher values of TPU in $T$ can be an indicator
of \organizedb.

\begin{equation} \label{eq:temporal-tpu}
\funcname{temporal-tpu\#}(T)= \frac{\displaystyle\sum_{t \in T}  1} { \mid \funcname{users}(T) \mid}
\end{equation}

\noindent\textbf{Retweeted tweet features}: 
Features based on retweets can be sign of bot account existence and \organizedb, because in case of retweets there is no need to
generate a content, which can be challenging for the automated accounts. 
The following are computed for the retweeted tweets :
\begin{description}
\item[Retweet frequency:]{The number of retweets ((i.e. if $100$ users have retweeted $t_x \in T$, this counts as $100$ retweets):

\begin{equation} 
\funcname{retweet\#}(T) = \displaystyle\sum_{t \in T} retweeted?(t) 
\end{equation}

\begin{equation} \label{eq:temp-retweet}
 \funcname{retweet\%}(T) = \frac{\funcname{retweet\#}(T) } { \mid T \mid}
\end{equation}

}

\item[Unique retweeted frequency:]{The percentage of distinct retweets among retweeted tweets.
It shows the diversity in the retweets. If most of the retweets are retweets of the same tweet, it can be a sign of a collective behavior.

\begin{equation} \label{eq:original-retweeted-tweet}
\funcname{original-retweeted-tweet\%}(T) = \frac{\Bigg\lvert \bigcup\limits_{\substack{t \in T\\retweeted?=1}} \{ t \} \Bigg\lvert} {\displaystyle\sum_{t \in T} retweeted?(t) }
\end{equation}
}

\item[Retweeting user frequency:]{Show how many of the users in the temporal tweet set are participated in retweets about \tracedHashtag. If in a temporal tweet set, most of the users are retweeted a tweet about \tracedHashtag, this can be a sign of collective behavior.

\begin{equation} 
\funcname{retweeted-users\#}(T) = \displaystyle\sum_{u \in users(T)}\ceil{\frac{\displaystyle\sum_{t \in tweets(u,T)} retweeted?(t)}{\mid tweets\#(u,T) \mid}}
\end{equation}

\begin{equation} \label{eq:retweeted-users}
\funcname{retweeted-users\%}(T) = \frac{\funcname{retweeted-users\#}(T)} { \mid \funcname{users}(T) \mid}
\end{equation}
}
\end{description}

\noindent\textbf{Features for tweets that are not retweeted:} These features focus on tweets and users who are not subject to retweeting. They are extracted because higher values in these features may suggest \organicb\ due to difficulty of creating an original tweet.

\begin{description}
	\item[Unretweeted tweet frequency:] {
		\begin{equation} \label{eq:unretweeted-tweets-percentage}
		\funcname{unretweeted\%}(T) = 1 - \funcname{retweet\%}(T)
		\end{equation}
	}
	\item[Users with no retweets frequency:]{
		\begin{equation} \label{eq:unretweeted-users-percentage}
		\funcname{unretweeted-users\%}(T) =   1 - \funcname{retweeted-users\%}(T)
		\end{equation}
	}

\item[Unretweeted tweet count:] {
	
	\begin{equation} \label{eq:unretweeted-tweets}
		 \funcname{unretweeted\#}(T) = \mid T \mid - \funcname{retweet\#}(T)
	\end{equation}
 }

\item[Count of users with no retweets frequency:]{
	
	\begin{equation} \label{eq:unretweeted-users}
		 \funcname{unretweeted-users\#}(T) =   \mid \funcname{users}(T) \mid - \funcname{retweeted-users\%}(T)
	\end{equation}
}

\item[Ratio of unretweets and users with no retweets:]{
	\begin{equation} \label{eq:unretweeted-users-ratio}
		\funcname{unretweeted-tweet_user_ratio}(T) =   \frac{unretweeted\#(T)}{unretweeted-users\#(T)}
	\end{equation}
}

\end{description}

\noindent\textbf{Mention Features:} 
Since there are usually groups who are pro and against the candidates in political domain, 
it is likely that users from different sides can have a dispute in Twitter. 
In these disputes, users use mentions to target other users.
Furthermore, in order to propagate messages, bot users may use mentions to get attention of other users.
Therefore, mention related features could be beneficial for \organizedb\ detection.

\begin{description}
	\item[Mention Ratio:]{Frequency of mentions to distinct mentions:
	\begin{equation} \label{eq:mention-ratio}
	\funcname{mention-ratio}(T) = \frac{\mid \bigcup\limits_{t \in T} mentions(t) \mid}	{\displaystyle\sum_{t \in T}  mention\#(t)}
	\end{equation}
	}

	\item[Ratio of mentions in retweets:]{Frequency of mentions that occur in retweeted tweets.
	\begin{equation} \label{eq:retweeted-mention-ratio}
	\funcname{mention-RT}(T) = \frac{\mid \bigcup\limits_{\substack{t \in T \\ \funcname{retweeted?}(t) = 1}} mentions(t) \mid}
	{\displaystyle\sum_{t \in T}  mention\#(t) \formulaspace \funcname{retweeted?}(t)}
	\end{equation}}
	\item[Ratio of mentions in unretweeted tweets:]{Frequency of mentions in tweets that are not retweeted.
	\begin{equation} \label{eq:unretweeted-mention-ratio}
	\funcname{mention-notRT}(T) =  \frac{\mid \bigcup\limits_{\substack{t \in T \\ \funcname{retweeted?}(t) = 0}} mentions(t) \mid}
	{\displaystyle\sum_{t \in T}  mention\#(t) - \displaystyle\sum_{t \in T}  mention\#(t) \funcname{retweeted?}(t)}
	\end{equation}
	}

\end{description}

\noindent\textbf{Sentiment Features} Within the tweets that contains a mention, the percentage of tweets with the sentiments \textit{VeryNegative}, \textit{Negative}, \textit{Neutral}, \textit{Positive}, \textit{VeryPositive} are computed.
The sentiment analysis of tweets are determined using the Stanford Core NLP tool~\cite{StanfordCoreNLP}.
In our study, tweet sentiments are only calculated for reply tweets because of the likelihood of disputes in political domain.
In these disputes, large amount of negative tweets directed to a user may be sign of \organizedb.

\begin{equation} \label{eq:mention-sentiment}
\funcname{mentioned-sentiment\%}(s,T) = \frac{\mid \{ t | t \in T, \funcname{sentiment}(t) = s \}\mid }{|T|}
\end{equation}

\subsection{Collection Representation}
\label{sec:featureSummarization}

To gain information about a collection its users and tweets (within a time interval) are characterized by numerous features extracted from data fetched from Twitter.
This feature extraction phase yields a large set of features representing each user and temporal analysis.
Any assessment about the collection requires the representation of the collection as a whole, which will be used in the training data set. 

\begin{figure}
	\begin{center}
		\includegraphics[width=0.6\paperwidth,height=0.25\paperheight]{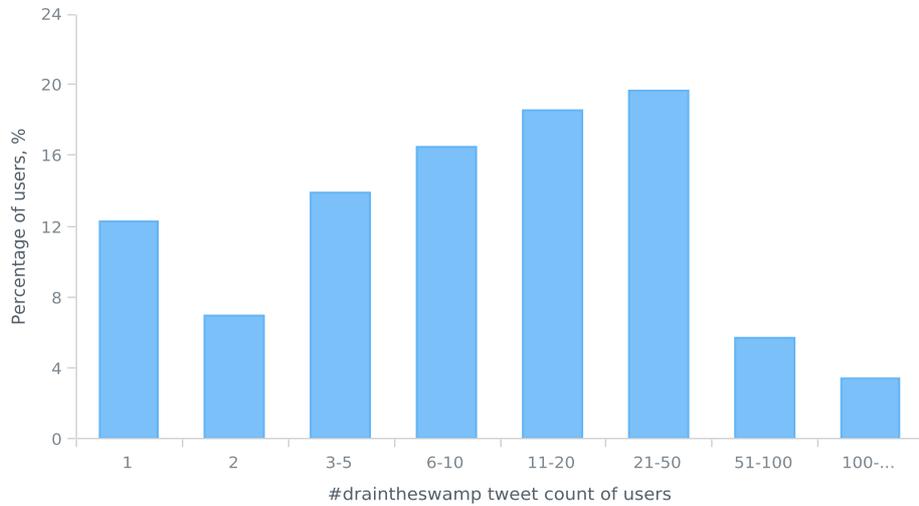}
	\end{center}
	\caption{Percentage of users based tweet counts with \#draintheswamp hashtag (Equation~\ref{eq:user-thashtag-use})}
	\vskip\baselineskip
	\label{fig:UserHashtagPostPercentage}
\end{figure}

Collections are uniformly represented by distributing the extracted values for each user feature into buckets defined for each feature.
Each feature value is placed in the bucket whose criteria it matches.
As an example, consider Figure~\ref{fig:UserHashtagPostPercentage}, which shows the hashtags posted by the users of the \#draintheswamp collection based on Equation~\ref{eq:user-thashtag-use}.
These values are summarized as follows: hashtagPostCount\_1~:~12.2\%, hashtagPostCount\_2~:~6.9\%,..., hashtagPostCount\_100-...~:~3.44\%.
Note that, the buckets are defined based on the observations from inspecting many collections.

In addition to the feature buckets, the mean, variance, standard deviation, minimum and maximum values are computed for each feature. All these values associated with a collection are stored in the Feature Repository~(Algorithm~\ref{alg:GeneralFeatureExtractionAlgorithm}).

\begin{figure}
	\begin{center}
		\includegraphics[width=0.6\paperwidth,height=0.25\paperheight]{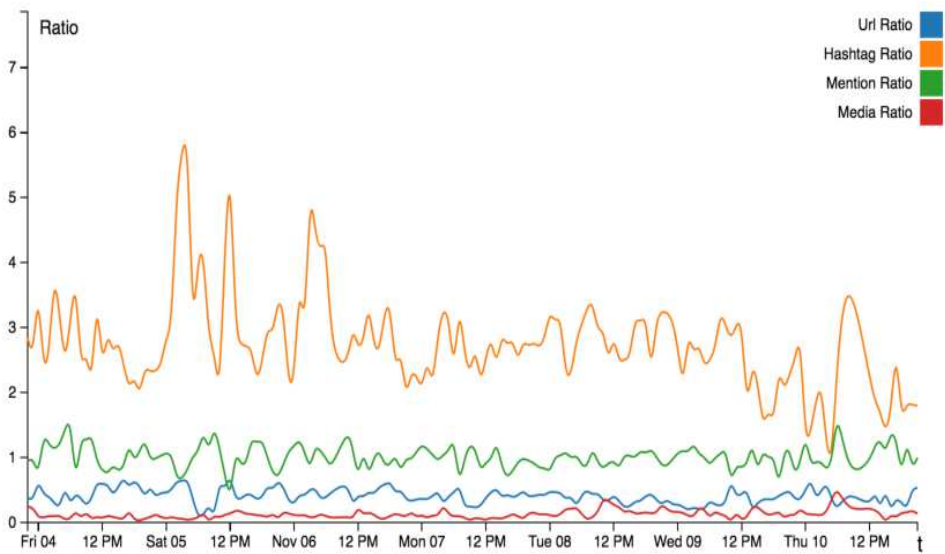}
	\end{center}
	\caption{Hourly entity ratios of \#draintheswamp hashtag(Equation~\ref{eq:EntityUseTI})}
	\vskip\baselineskip
	\label{fig:hourlyEntityRatio}
\end{figure}

\begin{figure}
	\begin{center}
		\includegraphics[width=0.65\paperwidth,height=0.25\paperheight]{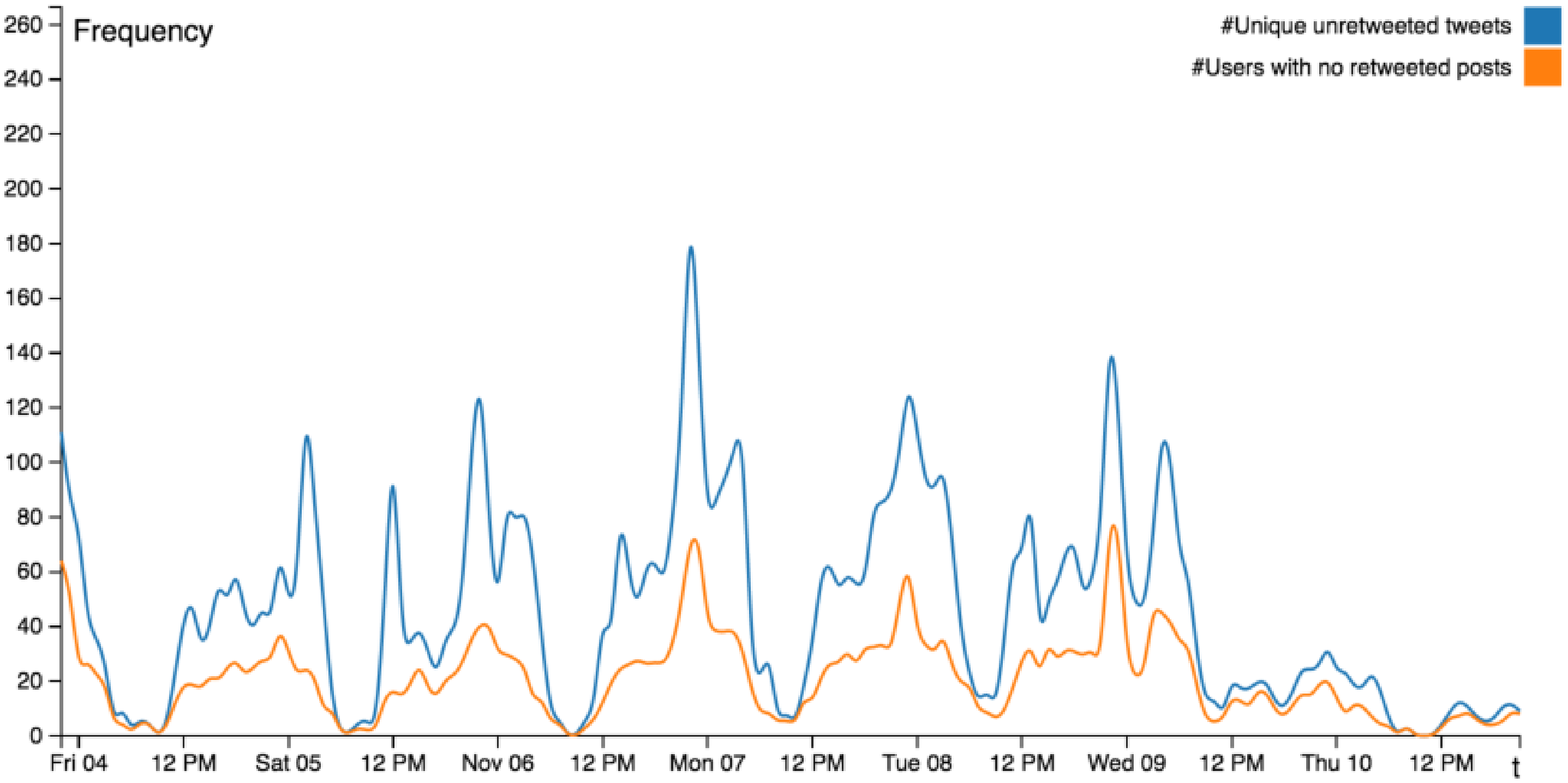}
	\end{center}
	\caption{Hourly Distribution of non-retweet count and distinct user count of \#draintheswamp hashtag(Equation~\ref{eq:unretweeted-tweets},\ref{eq:unretweeted-users} )}
	\vskip\baselineskip
	\label{fig:hourlyNonRetweetUser}
\end{figure}

On the other hand, in the summarization of temporal features, the values are not separated to the bucket values.
For each temporal interval feature, statistical values(mean, variance, standard deviation, min \& max values) are calculated.
Figure~\ref{fig:hourlyEntityRatio} and Figure~\ref{fig:hourlyNonRetweetUser} illustrate the extracted temporal features(Equation~\ref{eq:EntityUseTI} - Equation~\ref{eq:unretweeted-tweets},\ref{eq:unretweeted-users} ) of \#draintheswamp collection.

In the training data set, the resulting collection representation is used.  

\section{Experiments and Results}
\label{sec:experiment}

To assess the performance of the proposed approach,
the classifier models are tested on the training data set using 10-fold cross validation.
The results were promising for the \organizedb\ classification. This 
motivated us to proceed with the generation of  two more classifiers that use the same features for two different classification problems:
\textit{political} vs. \textit{non-political} and \textit{pro-Trump} vs \textit{pro-Hillary}. 
For each classification task, the training data set is labeled differently.

\subsection{Data: Tweet Sets}
\label{sec:tweetSets}

Training data sets are generated by using extracted features from expanded tweet sets(\textit{ET})
In the training data set a row is labeled with the label of the collection it is generated from.
Table~\ref{table:trainingDataSize} show the sizes of \ST\ (\#\ST), ET (\#ET),
the number of tweets used for temporal features extraction (\#TFT) and number of users(\#Users) for some collections \footnote{Sizes of the all collections in data set: \url{https://github.com/Meddre5911/DirenajToolkitService/blob/master/organizedBehaviorDataSets/OrganizedBehaviorDataSetSizes.csv}}\enlargethispage{-\baselineskip}. 

\begin{table}[h]
	\vskip\baselineskip 
	\caption[The number of tweets and users in various collections.]{\label{table:trainingDataSize} The number of tweets and users in various labeled collections. The expansion of the seed collections result in significant increases in number of tweets. }
	\begin{center}
		\begin{tabular}{|@{}l|r|r|r|r|l|}\hline			
			\multicolumn{1}{|c|}{\bfseries Hashtag} & \multicolumn{1}{c|}{\bfseries \#ST} & \multicolumn{1}{c|}{\bfseries \#ET} & \multicolumn{1}{c|}{\bfseries \#TFT} &\multicolumn{1}{c|}{\bfseries \#Users} & \multicolumn{1}{c|}{\bfseries Label}  \\\hline  
				\#imwithher& 35,362 & 4,411,703 & 75,681 & 15,792 & organized\\\hline
				\#maga &65,643 & 4,193,718 & 171,991 & 13,773 & organized\\\hline
				\#crookedhillary &18,999 & 3,773,531& 34,190 & 8,159 & organized\\\hline
				\#votetrump & 40,282 & 3,058,863 & 60,934 & 10,810 & organized\\\hline
				\#ladygaganewvideo &2,041& 310,908 &7,129 &1,486 & organized\\\hline
				\#news & 32,949 & 2,425,145 & 191,856 & 6,020 & organic\\\hline
				\#oscarfail &2,111 &749,748 & 2,152 & 1,507 & organic\\\hline
				\#thanksgiving & 4,575 & 539,081 & 4,461 & 2,367 & organic\\\hline
				\#brangelina & 2,288 & 347,596 & 2,961 & 1,676 & organic\\\hline
				\#unitedairlines & 1,345 & 276,228 & 2,240 & 898 & organic\\\hline
		\end{tabular}
	\end{center}
\end{table}

\subsubsection{Organized Tweet Sets}
\label{sec:organizedTweetSets}

To create a training data set, first \ST\ consisting of tweets with a hashtag must be gathered.
A hashtag is labeled as organized if the presence of \organizedb\ is observed within its tweets based on our manual inspection (i.e. \#LadyGagaNewVideo, \#podestaemails, \#podestaemails13, \#podestaemails20, \#BoycottHamilton).
Suitable hashtags reported the news or literature are also labeled organized (i.e. \#tcot \& \#pjnet based on a study\cite{NazisVsISISTwitter}).
Hashtags used by large amount of bots during the 2016 US election are also labeled as organized\footnote{In the first presidential debate, 32.7 percent of pro-Trump hashtags and 22.3 percent of pro-Hillary hashtags were posted by bots\cite{BotsAndAutomationOverTwitter}}. 
These hashtags are reported in studies regarding the prevalence of bot accounts during the 2016 US election~\cite{BotsAndAutomationOverTwitter,PresedentialElectionOnline,BotsAndAutomationOverTwitter2}. 
For externally discovered hashtags, if their bot based content generation is greater than 20\%, a collection is created for them (i.e. 
 \#AmericaFirst, \#MakeAmericaGreatAgain, \#TrumpPence16, \#NeverHillary, \#VoteTrump, \#CrookedHillary,  \#LoveTrumpsHate, \#NeverTrump, \#StrongerTogether, \#ImWithHer, \#UniteBlue.)

\begin{figure}
	\begin{center}
		\includegraphics[height= 0.25\paperheight]{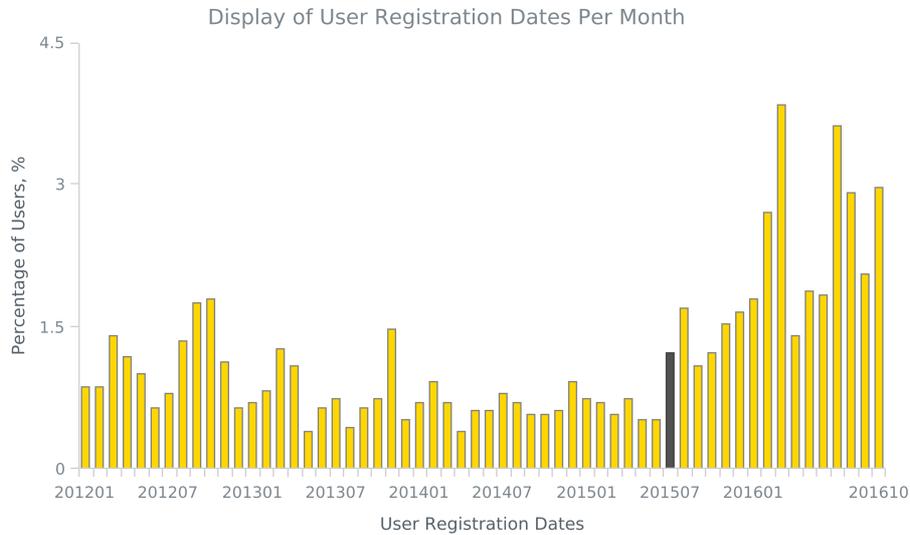}
	\end{center}
	\caption{Percentage of same users in Table~\ref{table:sameUserAnalysis} (users in \#benghazi \& \#crookedHillary) based on Twitter registration date}
	\vskip\baselineskip
	\label{fig:sameUserCreationDate}
\end{figure}

When seeking for candidate hashtags, we create a backlog. 
The ones that are not yet manually inspected are compared with collections known to be organized.
If there are significant  number of overlapping users, the hashtag is considered as organized.
This is due to the observation that colluding users are active over time and use several persisting hashtags to manipulate others. 
Table~\ref{table:sameUserAnalysis} shows the users who participated in multiple pro-Trump hashtags in different times. 
Furthermore, the percentage of users who registered after July 2015\footnote{The official nomination of Trump for presidency.} are in increasing trend~(Figure~\ref{fig:sameUserCreationDate}).
A suspected hashtag is labeled as organized if at least 20 percent of its users  participated in bot based collections.
Some of the tweet sets labeled in this manner are based on hashtags of \#benghazi, \#obamacarefail, \#imvotingbecause, \#draintheswamp, \#trumpwon, \#clintonemails, \#auditthevote, and \#hillaryemails.

In the tweet sets collected at different timess based on the bots cited in \cite{BotsAndAutomationOverTwitter} and \cite{PresedentialElectionOnline} numerous common hashtags were observed (i.e. \#election2016,\#tcot, \#imwithher, \#nevertrump, \#neverhillary, \#trumppence2016, \#p2).
This suggests that the use of some hashtags persist throughout a campaign, such as the observed in the US election 2016 campaign.
This is not surprising, since hashtags serve as group specific identifiers that serve to unite messages and people.

\begin{table}[h]
	\vskip\baselineskip 
	\caption[An example organized collection (\#benghazi) whose users are found in other trending collections.]{An example organized collection (\#benghazi) whose users are found in other trending collections.}
	\begin{center}
		\begin{tabular}{|l|l|l|c|c|}\hline		
			\textbf{Traced hashtag}& \multicolumn{4}{l|}{\#benghazi} \\\hline
			\textbf{Time Interval}& \multicolumn{4}{l|}{$24$ Oct - $1$ Nov $2016$}\\\hline
			\textbf{User\#}&  \multicolumn{4}{l|}{$7,854$}\\\hline\hline
			\multicolumn{5}{|c|}{Collections with mutual users with \#benghazi }\\\hline
			\textbf{Traced Hashtag} & \multicolumn{1}{c|}{\textbf{Interval}} & \textbf{User\#} & \multicolumn{1}{c|}{\textbf{$\bigcap$} (\#)} & \multicolumn{1}{c|}{\textbf{$\bigcap$} (\%)} \\\hline
			\#podestaemails & 17-24 Oct & $32,794$ &$3,232$ & $41.15$ \\\hline
			\#crookedhillary & 04-10 Nov  & $16,854$ & $2,291$ & $29.17$ \\\hline
			\#makeamericagreatagain & 04-10 Nov & $27,625$ & $2,311$ & $29.42$ \\\hline
			\#boycotthamilton & 17-22 Nov  & $36,561$ & $1,580$ & $20.12$ \\\hline
			\#maga & 20-28 Nov  & $29,130$ & $3,063$ &   $39.00$\\\hline	
		\end{tabular}
		\label{table:sameUserAnalysis}
	\end{center}
\end{table}

\subsubsection{Organic Tweet Sets }

Hashtags are labeled as organic when they are deemed to be spontaneously posted. 
In case of trending topics, such hashtags emerge due to events like holidays, natural disaster, and news about popular people.  
For example, the hashtags related the United Airlines, such as \#unitedairlines, \#boycottUnitedAirlines, \#newunitedairlinesmottos, instantly became popular after a video of passenger, who was forcibly removed from a plane due to over booking, went viral.
Similarly, during the 2017 Oscars ceremony, the \#oscarsfail hashtag became top trending after the best picture award was accidentally given to wrong movie.

Hashtags related to special days like \#Thanksgiving, \#LaborDay, \#NationalSiblingsDay, and \#WomansDay are also tagged as \textit{organic}, since these hashtags are used by a large number of diverse people.
Likewise, topic centric hashtags are widely used by diverse sets of users, such as \#news, \#deeplearning, and \#bigdata are labeled as organic.

However, recall that during the labelling of organized hashtags, we labelled hashtags as organized in case of considerable amount of
their contents are created by automated bots.
For the hashtags of \#Thanksgiving, \#LaborDay, \#NationalSiblingsDay, and \#WomansDay, 
it is also possible that they may contain bot activity, especially when they are trending topics.

In order to understand the level of bot existence in these organic hashtags, we examined characteristics of their tweets by calculating
statistical values of $DW(\%)$, $TPU(\mu)$, $RT(\%)$, $HT(\sigma^2)$, and $HT(\sigma)$ (the metrics used during the manual inspection of tweets in Section \ref{sec:approach}). As a result of these calculations, we conclude that these hashtags are used spontaneously.

For example, given the values for  \#womansDay shown in Table~\ref{table:manuelInspectionValues}, higher value of $DW(\%)$ indicates less similar messages, and lower value of $TPU(\mu)$ shows less multiple posts by the same users.
Likewise, lower values of $HT(\sigma^2)$ \& $HT(\sigma)$ indicate less hashtag use in \#womansDay tweets.

Furthermore, the overlapping users in organic hashtags inspected with the idea that if a user is present in numerous organic hashtags, it may be an incidence of bots hijacking trending hashtags.
Table~\ref{table:sameUserAnalysis} shows overlapping users in multiple
hashtags regarding special days, however their percentage is very low compared to those  in organized hashtags(see~Table~\ref{table:sameUserAnalysis}).

\begin{table}
	\vskip\baselineskip 
	\caption[An example organic collection (\#internationalwomensday) whose users are found in other trending collections.]
	{\label{table:organicSameUserAnalysis}An example organic collection (\#internationalwomensday) whose users are found in other trending collections.}
	\begin{center}
		\begin{tabular}{|@{}l|r|r|r|c|}\hline		
			\textbf{Traced hashtag}& \multicolumn{4}{l|}{\#internationalwomensday} \\\hline
			\textbf{Time Interval}& \multicolumn{4}{l|}{$07-09$ March $2017$}\\\hline
			\textbf{User\#}&  \multicolumn{4}{l|}{$267,695$}\\\hline\hline
			\multicolumn{5}{|c|}{Collections with mutual users with \#internationalwomensday}\\\hline
			\textbf{Traced Hashtag} & \multicolumn{1}{c|}{\textbf{Interval}}  & \textbf{User\#} &\multicolumn{1}{c|}{\textbf{$\bigcap$} (\#)} & \multicolumn{1}{c|}{\textbf{$\bigcap$} (\%)} \\\hline
			\#Thanksgiving &20-28.Nov.16 & $104,060$ & $6,247$ & $6$\\\hline
			\#Oscars &25.Feb-01.Mar.17 & $134,868$ & $12,656$ & $9.38$\\\hline
			\#NationalPetDay & 09-14.Apr.2017 & 39,506 & $3,095$ & $7.83$\\\hline
			\#NationalSiblingsDay & 09-14.Apr.2017 & $58,908$  & $4,220$ & $7.16$\\\hline
			\#WorkersDay & 30.Apr-02.May.2017 & $1,269$ & $71$ & $5.59$\\\hline
			\#MothersDay & 13-15.May.2017 & $113,225$ & $8,207$ & $7.25$\\\hline
		\end{tabular}	
	\end{center}
\end{table}

\subsubsection{Political vs. Non-Political}

The labeling of political and non-political hashtags are done by manual inspecting  \ST.
The data set labeled as political is created from 2016 US election data.
The determination of a label was fairly straightforward in comparison to organized behavior.  

\subsubsection{Pro-Trump, Pro-Hillary, Neither}

Specifically in political campaigns, there are strong groups who are pro and against certain candidates.
Meticulous attention is given to reaching people and delivering propaganda that supports their candidate and damages other candidates. 
In the 2016 US election the political campaigns were fierce and quite different than earlier campaigns.
Heavy use of social media was deployed from within the country as well as from other countries.
Just what happened during the campaign and elections has been a topic of heated debate that is unfolding at the time of writing this paper.

The hashtags are mainly selected and labeled based on given \textit{pro-Trump} and \textit{pro-Hillary} hashtags in the work on bots and automation in Twitter\cite{BotsAndAutomationOverTwitter},  
The remaining hashtags are manually labeled as usual (with use of Table~\ref{table:sameUserAnalysis} and Table~\ref{table:organicSameUserAnalysis}), except for hashtags that do not belong to either class that are labeled as \textit{None}.

\subsection{Training  Data Sets}
\label{sec:experimentTrainingDataSets}

The labels of training data sets for each category are as follows\footnote{Training data sets can be found in \url{https://github.com/Meddre5911/DirenajToolkitService/tree/master/organizedBehaviorDataSets/trainingDataSets}}:

\begin{description}
	\item [organic vs organized]: $851$ records of which $625$ are labeled as organized and $226$ organic.
	\item [political vs non-political]: $879$ records of which $231$ are labeled as non-political and $648$ political.
	\item [pro-Trump vs pro-Hillary]: $854$ records of which $311$ are labeled as pro-Trump, $171$ pro-Hillary, and $371$ None. 
\end{description}

For each category, four different training data sets are generated based on the features of all collections:
(1) all extracted features, (2) principal components of all extracted features obtained from PCA(see~Section~\ref{sec:ClassificationMethods}), (3) all extracted features except user features based on \tracedHashtag\ (see~ Equations~\ref{eq:user-thashtag-use},\ref{eq:daily-average-tweet},\ref{eq:user-daily-tweet-comparison}), and (4) principal components of the features in (3).

Data sets (3) and (4) ignore \tracedHashtag\ related user features for the purpose of assessing how the model behaves when the features of the hashtag that were used to fetch the initial data are eliminated. 
Since hashtags are not the only mechanism for coordination on Twitter, one would expect the method to remain successful. 
By removing the \tracedHashtag\ related features in Equations~\ref{eq:user-thashtag-use},\ref{eq:daily-average-tweet},\ref{eq:user-daily-tweet-comparison}, we can see how reliant the detection is to hashtags.

In the second and fourth data sets, PCA is used in order to examine impact of dimension reduction on the  performance of the classification, with consideration of the need for real-time processing.
Table~\ref{table:trainingDataSize} provides information about the size  of some collections in the collection repository.
Even though the size of collections are highly limited by Twitter Search API, (\#ET) sizes are in the millions.
  
\subsection{Results of three Classifers}
\label{sec:experiments-results}

The model is evaluated for each classifier and data set, using 10-fold cross validation. 
The organic vs organized and political vs non-political are binary classification models, 
whereas the pro-Trump vs pro-Hillary vs None classification is a multi-classification problem. 

Results show that the random forest algorithm results in high scores with full features,  while logistic regression and SVM algorithms give better results when PCA is applied.

Furthermore, the performance with Data~Set~3~and~4 shows that the presence or absence of \tracedHashtag\ is not essential to capture organized behavior.
Tables~\ref{table:EvalOrganicOrganized}, \ref{table:PoliticalNonPolitical}, and \ref{table:proTrumpHillary} show the results for each category. 
SVM results are not provided for pro-Trump vs pro-Hillary classification due to lack of support for multi-classification with SVM in Spark MlLib\cite{sparkMlLibguide}.

\begin{table}[]
	\centering
	\caption{Evaluation of alternative approaches for classifying organic vs. organized collections. F is F-Measure, A is Accuracy, and ROC is Receiver Operating Characteristic.}
	\label{table:EvalOrganicOrganized}
	\begin{tabular}{|c|l|c|c|c|}
		\hline
		\textbf{Features}        & \textbf{Method}&  \textbf{A} &  \textbf{F} & 	\textbf{ROC}  \\ 
		\hline
		\multicolumn{1}{|c|}{All} & \textbf{Random Forest} & $0.99$   & $0.99$    & $0.99$  \\ \cline{2-5} 
		\multicolumn{1}{|c|}{Features}                       & \textbf{Logistic Regression} & $0.99$  & $0.99$  & $0.99$  \\ \cline{2-5} 
		\multicolumn{1}{|c|}{}                       & \textbf{SVM}  & $0.75$ & $0.64$ & $0.66$ \\ \hline
		
		\multicolumn{1}{|c|}{PCA} & \textbf{Random Forest} & $0.98$   & $0.97$    & $0.96$  \\ \cline{2-5} 
		\multicolumn{1}{|c|}{of all}                       & \textbf{Logistic Regression} & $0.98$  & $0.98$  & $0.98$  \\ \cline{2-5} 
		\multicolumn{1}{|c|}{Features}                       & \textbf{SVM}  & $0.99$ & $0.99$ & $1.00$ \\ \hline
		
		\multicolumn{1}{|c|}{All except} & \textbf{Random Forest} & $0.99$   & $0.99$    & $0.99$  \\ \cline{2-5} 
		\multicolumn{1}{|c|}{\tracedHashtag} & \textbf{Logistic Regression} & $0.99$  & $0.98$  & $0.97$  \\ \cline{2-5} 
		\multicolumn{1}{|c|}{Features}                       & \textbf{SVM}  & $0.75$ & $0.64$ & $0.66$ \\ \hline
		
		\multicolumn{1}{|c|}{PCA of all} & \textbf{Random Forest} & $0.97$   & $0.96$    & $0.95$  \\ \cline{2-5} 
		\multicolumn{1}{|c|}{except \tracedHashtag}                       & \textbf{Logistic Regression} & $0.98$  & $0.98$  & $0.97$  \\ \cline{2-5} 
		\multicolumn{1}{|c|}{Features}                       & \textbf{SVM}  & $0.99$ & $0.99$ & $0.99$ \\ \hline
	
	\end{tabular}
\end{table}
\begin{table}[]
	\centering
		\caption{Results of alternative approaches to classifying Political vs. Non-Political collections. F is F-Measure, A is Accuracy, and ROC is Receiver Operating Characteristic.}
	\label{table:PoliticalNonPolitical}
	\begin{tabular}{|c|l|c|c|c|}
		\hline
		\textbf{Features}        & \textbf{Method} &  \textbf{A} &  \textbf{F} & \textbf{ROC}  \\ \hline
		\multicolumn{1}{|c|}{All} & \textbf{Random Forest} & $0.99$  & $0.99$   & $0.99$  \\ \cline{2-5} 
		\multicolumn{1}{|c|}{Features}                       & \textbf{Logistic Regression} & $0.99$  & $0.99$  & $0.99$  \\ \cline{2-5} 
		\multicolumn{1}{|c|}{}                       & \textbf{SVM}  & $0.77$& $0.65$& $0.64$ \\ \hline
		
		\multicolumn{1}{|c|}{PCA} & \textbf{Random Forest} & $0.99$   & $0.98$   & $0.97$ \\ \cline{2-5} 
		\multicolumn{1}{|c|}{of all}                       & \textbf{Logistic Regression} & $0.98$  & $0.98$ & $0.98$  \\ \cline{2-5} 
		\multicolumn{1}{|c|}{Features}                       & \textbf{SVM}  & $1.00$ & $0.99$ & $1.00$\\ \hline
		
		\multicolumn{1}{|c|}{All except} & \textbf{Random Forest} & $0.99$  & $0.99$    & $ 0.99$ \\ \cline{2-5} 
		\multicolumn{1}{|c|}{\tracedHashtag}                       & \textbf{Logistic Regression} & $0.99$ & $0.99$  & $0.99$  \\ \cline{2-5} 
		\multicolumn{1}{|c|}{Features}                       & \textbf{SVM}  & $0.77$& $0.65$& $0.64$ \\ \hline
		
		\multicolumn{1}{|c|}{PCA of all} & \textbf{Random Forest} & $0.98$  & $0.98$  & $0.98$ \\ \cline{2-5} 
		\multicolumn{1}{|c|}{except \tracedHashtag}                       & \textbf{Logistic Regression} & $0.99$  & $0.98$  & $0.99$  \\ \cline{2-5} 
		\multicolumn{1}{|c|}{Features}                       & \textbf{SVM}  & $1.00$ & $0.99$& $1.00$ \\ \hline
	\end{tabular}
\end{table}

\begin{table}[]
	\centering
	\caption{Results of alternative approaches for classification of Pro-Hillary vs. Pro-Trump vs None collections. A is accuracy, F is F-measure, P is precision, and R is recall. }
	\label{table:proTrumpHillary}
	\begin{tabular}{|c|l|c|c|c|c|}
	\hline
		\textbf{Features}        &  \textbf{Method} &  \textbf{A} &  \textbf{F} & 	\textbf{P} & \textbf{R} \\ 
        \hline
		\multicolumn{1}{|c|}{All} & \textbf{Random Forest} & $0.97$  & $0.96$   & $0.97$ & $0.96$  \\ \cline{2-6} 
		\multicolumn{1}{|c|}{Features}                       & \textbf{Logistic Regression} & $0.44$  & $0.20$  & $0.15$ & 0.33  \\ \cline{2-6} 
		\hline
		\multicolumn{1}{|c|}{PCA of all} & \textbf{Random Forest} & $0.92$   & $0.90$   & $0.93$ & $0.88$ \\ \cline{2-6} 
		\multicolumn{1}{|c|}{Features}                       & \textbf{Logistic Regression} & $0.94$  & $0.93$ & $0.93$ & $0.93$ \\ \cline{2-6} 
		\hline
		\multicolumn{1}{|c|}{All except} & \textbf{Random Forest} & $0.96$  & $0.95$    & $0.96$ & $0.94$ \\ \cline{2-6} 
		\multicolumn{1}{|c|}{\tracedHashtag Features}                       & \textbf{Logistic Regression} & $0.44$ & $0.20$  & $0.15$ & $0.33$  \\ \cline{2-6} 
		\hline
		\multicolumn{1}{|c|}{PCA of all except} & \textbf{Random Forest} & $0.92$  & $0.89$  & $0.93$ & $0.87$ \\ \cline{2-6} 
		\multicolumn{1}{|c|}{\tracedHashtag\ Features}                       & \textbf{Logistic Regression} & $ 0.93$  & $0.91$  & $0.92$ & $0.91$  \\ \cline{2-6} 
		\hline
	\end{tabular}
\end{table}

\subsection{Top five features of classifications}
\label{sec:topFiveFeatures}

Feature selection is important for improving accuracy and decreasing training time.
The following sections enumerate the top five features\footnote{Features are selected using the \emph{ClassifierSubsetEval} attribute evaluator with the Random Forest classifier and the Best First search method in the Weka\cite{wekaDefinitions} by applying 10 fold cross validation} for each classification. 

References to related functions and feature equations are provided next to the features.
$C$ refers to a collection and $U = users(C)$. 
All features assume $u \in U$, additional constraints are expressed when relevant.
Percentages, frequencies, maximum, minimum, $\sigma^2$, and $\sigma$ values are computed over $t \in C$ and $u \in tweets(u,C)$. 
For details regarding the functions and features see Section~\ref{sec:features}.
The most important five features during classifications for each data set are :
\begin{description}
\item{\textbf{Organic vs. Organized}}
\begin{description}
\item{All features:}
\begin{enumerate}
	\item $\sigma^2$ of $\funcname{media-use}(u,C)$ [Eq.~\ref{eq:userEntityUse}]
	\item $\%$ of users such that $\funcname{follower-degree}(u) > 1$ [Eq.~\ref{eq:followerDegree}]
	\item $\%$ of users such that $1 \leq favorite\#(u) \leq 100$  
	\item Minimum of $tweet\#(u)$
	\item Maximum $\funcname{tweet-hashtag\#}_{\mu/d}(u,D,C)$ [Eq.~\ref{eq:user-daily-tweet-comparison}
\end{enumerate}
\item{All features excluding the hashtags:}
\begin{enumerate}
	\item $\#$ of users such that $0.6 \leq \funcname{media-use}(u,C) \leq 0.9$ [Eq.~\ref{eq:userMediaUse}]
	\item $\#$ of users such that  $\funcname{hashtag-use}(u,C)=7$ [Eq.~\ref{eq:userHashtagUse}]
	\item $\#$ of users such that   $\funcname{mention-use}(u,C)=10$ [Eq.~\ref{eq:userMentionUse}]
	\item Minimum of $\funcname{follower-degree}(u)$ [Eq.~\ref{eq:followerDegree}]
	\item $\#$ of users such that $1 \leq favorite\#(u) \leq 100$ 
\end{enumerate}
\end{description}
\item{\textbf{Political vs. Non-Political classification results}}
\begin{description}
\item{All features:}
\begin{enumerate}
	\item $\sigma^2$ of  $\funcname{media-use}(u,C)$ [Eq.~\ref{eq:userMediaUse}]
	\item Maximum of $\funcname{mention-use}(u,C)$ of all users [Eq.~\ref{eq:userMentionUse}]
	\item $\mu$ of $\funcname{follower-degree}(u)$ [Eq.~\ref{eq:followerDegree}]
	\item $\mu$ of $10.001 \leq favorite\#(u) \leq 20.000$ 
	\item Maximum of $tweet\#(u)$ 
\end{enumerate}
\item{All features excluding the hashtags:}
\begin{enumerate}
	\item $\sigma^2$ of   $\funcname{media-use}(u,C)$ [Eq.~\ref{eq:userMediaUse}]
	\item $\mu$ of $\funcname{follower-degree}(u)$ [Eq.~\ref{eq:followerDegree}]
	\item $\%$ of users such that  $10.001 \leq favorite\#(u)  \leq 20.000$ 
	\item $\%$ of users such that  $10.001 \leq tweet\#(u) \leq 20.000$ 
	\item Maximum of $tweet\#(u)$ 
\end{enumerate}
\end{description}

\item{\textbf{pro-Hillary vs. pro-Trump vs. None}}
\begin{description}
\item{All features:}
\begin{enumerate}
	\item $\%$ of users such that $0.0001 \leq \funcname{media-use}(u,C) \leq 0.5$ [Eq.~\ref{eq:userEntityUse}]
	\item $\%$ of users such that  $\funcname{url-use}(u,C)=1$  [Eq.~\ref{eq:userURLUse}]
	\item $\%$ of users such that  $\funcname{mention-use}(u,C)=7$ [Eq.~\ref{eq:userMentionUse})
	\item$\sigma^2$ of  $\funcname{url-use}(u,C)$ [Eq.~\ref{eq:userURLUse}]
	\item $\%$ of users such that $\frac{\funcname{mention-use}(u,C)}{\funcname{tweet}\#_{\mu/d}(u)}=1$ [Eq.~\ref{eq:user-daily-tweet-comparison} \& Section~\ref{sec:featureSummarization}]
\end{enumerate}
\item{All features excluding the hashtags:}
\begin{enumerate}
	\item $\%$ of users such that $\funcname{mention-use}(u,C)=0$ [Eq.~\ref{eq:userMentionUse}]
	\item $\%$ of users such that $0.0001 \leq \funcname{media-use}(u,C)= \leq 0.5$ [Eq.~\ref{eq:userHashtagUse}]
	\item $\%$ of users such that $0.0001 \leq \funcname{mention-use}(u,C) \leq 0.5$ [Eq.~\ref{eq:userMentionUse}]
	\item $\%$ of users such that $\funcname{url-use}(u,C)=1$ [Eq.~\ref{eq:userEntityUse}]
	\item $\%$ of users such that $1 \leq favorite\#(u) \leq 100$ 
\end{enumerate}
\end{description}
\end{description}

It is surprising that none of the temporal features are in top five features, 
suggesting that  \organizedb\ detection could be achieved by extracting user features only.
One of the most common advise for increasing engagement is the timing of posts to increase likelihood of visibility. 
The top user-based are also reported in other studies\cite{DetectingSpamURLCao,OrganicorOrganizedUrlCao}, which is encouraging.
Compared to temporal features, extracting user features is easier and more efficient.
A classifier based only on user features might be useful.

The use of media stands out as an important factor, appearing in the top 5 of every classification.
This is not surprising, since it is generally known that tweets with media are viewed the most.
Bots and trolls use regularly use them. 
The consistent use of media entities can be observed from the variance.
Furthermore, consistent use of media results in high media ratio for the user (Equation~\ref{eq:followerDegree}), which may
signal bots or users in some organized behavior. 

A follower degree of $1$ means that the user does not follow anyone. 
To be effective, organized users are expected to follow users.
Therefore, a follower degree value $1$ would likely identify an organic collection.
On the contrary, a follower degree of $0$ implies not being popular -- having $0$ followers.
Newly deployed bots often have $0$ followers~\cite{botNews4}. 

Furthermore, a user's tweet\# and favorite\# are helpful in classification. 
When these values are in the thousands they are more likely to be associated to an automated user, whereas low values indicate regular users.

Very high URL, hashtag, and mention ratios imply the occurrence of multiple such entities in most of their tweets (Equation~\ref{eq:userEntityUse}).
Dedicating the small space provided by a tweet to deliver several entities suggests an intent to become more visible.
Mentions trigger notification to the mentioned and it may also noticed by the follower of the mentioned person. 
URLs extend the information that can be delivered by luring a user to an external site.
Finally, a high mention rate points to interaction with other users.

The daily hashtag rate (Eq.~\ref{eq:user-daily-tweet-comparison})
shows the daily use of the \tracedHashtag\ during the days that hashtag is used in Twitter.
A higher frequency of tweets than typical on a given day compounded by the occurence of the \tracedHashtag\ suggests an \organizedb.

\section{Discussion}
\label{sec:discussion}

The results presented in Section~\ref{sec:experiment} are promising for detecting various kinds of organized behavior. 
There are numerous directions to pursue in terms of extending the model, implementation techniques, and data collection. 

The three classification experiments (organized vs organic; political vs non-political; and pro-Trump vs pro-Hillary vs None) are somewhat different, focusing on the presence of any, topic-based, and group-based organization respectively.
All three models are trained with the same features and all yield promising results.  
Based on these results we speculate that the proposed features may fingerprint tweet collections, at least in terms of their nature. 

It could be that our data set is too small 
even though it contains millions of tweets\footnote{For details see: \url{https://github.com/Meddre5911/DirenajToolkitService/blob/master/organizedBehaviorDataSets/OrganizedBehaviorDataSetSizes.csv}}\enlargethispage{-\baselineskip}.
On the other hand, user features may indeed be most significant ones by shining light on the the characteristic of those who are in collusion.
A  classifier model generated solely based on user features may be worth developing, since it is fairly easy and cost effective, which would be beneficial for real-time classification.

Collection labeling is performed by manual inspection, which is enormously time consuming and  error prone. 
While the visualization tools that we created to summarize collections help, the task remains difficult. 
We spent endless hours inspecting thousands of tweets and in deliberation.
Better approaches for  labeling  training data sets should be explored.

The approach to constructing and processing collections must be systematically investigated to study the impact of various choices.
While the use of hashtags is simple and powerful, they are not the only mechanism for coordination and organization on Twitter.  
During our experiments, by using data sets, which ignore \tracedHashtag\ related features~(Section~\ref{sec:experimentTrainingDataSets}), we aimed to test performance of our approach
in case of tweets are collected based on other mechanisms such as community detection\cite{ozer2016community,bakillah2015geo} algorithms or topic detection\cite{topicDetection:YildirimUskudarli} methods.

The aim of the present work was to develop a basic model for detecting patterns of organized behavior on Twitter.
Limited resources also played a role in keeping it simple, however, 
it is interesting to learn how simple approaches perform.
We examined many features that are not in the current model that are related to tweet content, 
users relations, and information flow patterns.
We observed the presence of unusually similar content posted by the same persons or those connected to them during the same interval, presumably serving a shared agenda.
Similarity computation among all tweets in a set is very costly (complexity of $\mathcal{O}(n^2)$)-- a task that exhausted our resources. 
Efficient approaches to compare large sets of posts is an interesting research direction.

Tightly connected communities can be very effective in propagating information.
In popular context, such as political campaigns, follow relations exceeding 50K are not uncommon.
When the followers of followers of thousands of people are considered, the computation of closeness centrality becomes challenge. 
Limiting the number of hops to 2 nodes, the users stored in a Neo4j\footnote{Neo4j version:2.1.2} graph database became overwhelming.
Alternative approaches to address closeness centrality is an interesting direction.

Another observation is information flow patterns, for example recurring message paths, such as $A$ tweets $t_x$, which is retweeted by $B$, which is retweeted by $C$. Here the users $A,B,C$ remain the same, wheras the message $t_i$ may vary.
Such pattern suggests presence of coordination, quite possibly automated.

Although, the current model does not include  many features we examined, they remain of great interest.
The features we did include are those we considered to be significant and whose computation was in the realm of our resources.
Unfortunately, tweet sentiment detection, which is a highly explored area, has limited success due to the nature of tweets being short, messy, unconventional, and satirical. 
The performance of Stanford NLP tool\cite{manning-EtAl:2014:P14-5} used in this work is insufficient as of yet.

A great deal of effort went into developing the feature extraction system to produce the training data set. 
Many performance issues were experienced related to processing big data. 
Big Data systems used for real-time detection require high memory and high processing power (GPU) to run complex algorithms on very large data. 
To prevent excessive database I/O, in memory graph databases needs to be used. 
Big data frameworks and tools are appropriate for social media data, which is vast and created in a continuous flow. 

Our prototype implementation uses MongoDB. 
As the table sizes and I/O requests increased the performance of queries dramatically decreases (in spite of index use). 
To prevent bottlenecks in one table, temporary tables were created with indexes to reduce the wait-time.
Unfortunately, I/O bottlenecks persisted. 
For real time decision and detection systems there is a need for tools like Apache Spark\cite{Zaharia:2016:ASU:3013530.2934664}, which supports terabyte-scale in-memory data processing. 
Our repository is more than $480$ gigabytes with over $200$ million tweets, summary tables, indexes.
A Spark cluster can process all that data in memory.
Also, the Spark GraphX API would increase the feature extraction performance, especially for graph based features.

Our observations have strengthened our belief in how relevant it is to detect misinformation and manipulation.
With manipulators on social media getting more sophisticated, we expect that the detection of their activities will get more difficult.

To summarize, the results of this work are encouraging with many future directions to pursue.
The main future directions are with respect to validation, improved features,  performance, and domain specific detections.

\section{Conclusions}
\label{sec:futureworkandconclusions}

This work proposes a supervised learning based model for automatically classifying tweet sets that exhibits organized behavior patterns. 
Towards this end models are trained with user and temporal features extracted from over $200$ million tweets that were mostly gathered during the 2016 US presidential election.
 
Three types of classifications were performed among the categories: [organic,organized], [political, non-political], and [pro-Trump,pro-Hillary,None]. 
In each case, the random forest algorithm  consistently resulted in high accuracy and f-measure scores with an average of $0.95$. 

The results of classifying tweet sets  suggest that neither content nor user relation features are required to successfully classify them.
Furthermore, that user features are the most significant regarding our classification tasks.
Further investigated with larger training data sets should be perform to further validate.
Features like the closeness centrality and  tweet similarity are very costly, therefore it is encouraging that the detection of organized behavior can be achieved without them.
Organized vs organic behavior classification is well suited for politics and campaigns, as the organizing efforts carried out in Twitter are well captured in the model and features.
The results are encouraging and we intend to pursue this work and its domain specific applications.

\section{Acknowledgment}
\label{sec:acknowledgment}

This work is partially supported by the Turkish State Planning Organization (DPT) under the TAM Project, number 2007K120610. We thank the members of the Complex Systems Lab (SosLab) at the Computer Engineering Department of Bogazici Universitysupport for their support and constructive comments. 

\bibliographystyle{achemso}
\bibliography{references}

\end{document}